\documentclass[%
superscriptaddress,
amsmath,amssymb,
aps,
prc,
twocolumn,
floatfix
]{revtex4-1}

\usepackage{graphicx}% Include figure files
\usepackage{dcolumn}% Align table columns on decimal point
\usepackage{bm}% bold math
\usepackage{amsfonts}
\usepackage{hyperref}% add hypertext capabilities
\usepackage[mathlines]{lineno}% Enable numbering of text and display math
\usepackage{enumerate}

\usepackage{colortbl}
\newcommand{\snn}{\sqrt{s_\text{NN}}}

\begin{document}

%\preprint{}

\title{Search for the QCD Critical Point by Transverse Velocity Dependence of Anti-deuteron to Deuteron Ratio}

\author{Ning Yu}
\affiliation{School of Physics \& Electronic Engineering, Xinyang Normal University, Xinyang 464000, China}
\author{Dingwei Zhang}
\affiliation{Key Laboratory of Quark \& Lepton Physics (MOE) and Institute of Particle Physics, \\Central China Normal University, Wuhan, 430079, China.}
\author{Xiaofeng Luo}
\thanks{xfluo@mail.ccnu.edu.cn}
\affiliation{Key Laboratory of Quark \& Lepton Physics (MOE) and Institute of Particle Physics, \\Central China Normal University, Wuhan, 430079, China.}

\date{\today}% It is always \today, today,

\begin{abstract}
We propose the transverse velocity ($\beta_T$) dependence of the anti-deuteron to deuteron ratio as a new observable to search for the QCD critical point in heavy-ion collisions. The QCD critical point can attract the system evolution trajectory in the QCD phase diagram, which is known as focusing effect. To quantify this effect, we employ thermal model and hadronic transport model to simulate the dynamical particle emission along a hypothetical focusing trajectory near critical point. We found the focusing effect can lead to anomalous $\beta_T$ dependence of $\bar{p}/p$, $\bar{d}/d$ and $^3\overline{\text{He}}/^3\text{He}$ ratios. We examined the $\beta_T$ dependence of $\bar{p}/p$ and $\bar{d}/d$ ratios of central Au+Au collisions at $\snn = $ 7.7 to 200 GeV measured by the STAR experiment at RHIC. Surprisingly, we only observe a negative slope in $\beta_T$ dependence of $\bar{d}/d$ ratio at $\snn = $ 19.6 GeV, which indicates the trajectory evolution has passed through the critical region. In the future, we could constrain the location of the critical point and/or width of the critical region by making precise measurements on the $\beta_T$ dependence of $\bar{d}/d$ ratio at different energies and rapidity.
\begin{description}\item[PACS numbers]
\verb+25.75.Nq, 24.10.Lx, 24.10.Pa+
\end{description}
\end{abstract}

\maketitle
\section{Introduction}
Quantum Chromodynamics (QCD) is the fundamental theory of the strong interaction. One of the main goals of relativistic heavy-ion collisions is to explore the phase structure of the hot and dense QCD matter, which can be displayed in the $T-\mu_B$ plane ($T$: temperature, $\mu_B$: baryon chemical potential)  of QCD phase diagram. Lattice QCD calculations confirmed that the transition between hadronic gas and Quark-Gluon Plasma (QGP) is a smooth crossover at $\mu_B$=0~\cite{Aoki:2006we}. At large $\mu_B$ region, QCD based models predicted that the phase transition is of the first order~\cite{Fischer:2018sdj,Qin:2010nq,Shi:2014zpa,Lu:2015naa,Gao:2016qkh}. The QCD critical point (QCP) is the end point of the first order phase transitions boundary. Theoretically, many efforts have been made to locate the critical point in Lattice QCD ~\cite{FODOR200373,1126-6708-2004-04-050,PhysRevD.71.114014,Karsch:2015nqx,Gupta:2011wh} and models~\cite{Stephanov:2007fk}, but its position and even the existence is still not confirmed yet. Therefore, from the experimental side, scientists are performing a systematical  exploration of the phase structure of the QCD matter at high baryon density region. The search for the critical point is one of the main goals of the Beam Energy Scan (BES) program at the Relativistic Heavy-ion Collider (RHIC).  It is also the main physics motivation for future accelerators, such as Facility for Anti-Proton and Ion Research (FAIR) in Darmstadt and Nuclotron‐based Ion Collider fAcility (NICA) in Dubna. Experimental confirmation of the existence of the QCD critical point will be a milestone of exploring the nature of the QCD phase structure.

In the vicinity of the QCP, the correlation length of the system and density fluctuations will become large. In the first phase of Beam Energy Scan at RHIC (BES-I, 2010-2014), the STAR experiment has made two important measurements, which are dedicated to search for the QCP: 1). The measurement of the cumulants of net-proton,  net-charge and net-kaon multiplicity distribution~\cite{PhysRevLett.105.022302,PhysRevLett.112.032302,PhysRevLett.113.092301,2018551,Luo:2015ewa,Luo:2015doi} in Au+Au collisions at  $\snn = $7.7-200 GeV. One of the most striking findings is the observation of non-monotonic energy dependence of the fourth order net-proton cumulant ratios ($C_4/C_2$) in the most central (0-5\%) Au+Au collisions. We observe a minimum dip around 19.6 GeV and large increasing at 7.7 GeV. The review of these results can be found in ref.~\cite{Luo:2017faz}. 2). The measurement of the light nuclei (deuteron and triton) production as well as derived neutron density fluctuations at RHIC. We observe a non-monotonic energy dependence of the neutron density fluctuations in central (0-10\%) Au+Au collisions with a maximum peak around 19.6 GeV~\cite{Zhang:2019wun,Liu:2019nii}. These non-monotonic behaviors, the dip and peak structures observed around 19.6 GeV,  are qualitatively consistent with the theoretical predictions of the signature of the critical point~\cite{Stephanov:2011pb,Sun:2017xrx,CHEN20181}. 

It was predicted that the QCD critical point will serve as an attractor of the trajectory evolution in the $T-\mu_B$ plane, which is known as the QCP focusing effect~\cite{PhysRevC.71.044904,PhysRevLett.101.122302}. The entropy over baryon density ratio $s/n_b$ is constant along the isentropic trajectory. When the isentropic trajectory passes through the critical region in the $T-\mu_B$ plane, the transverse velocity ($\beta_T=p_T/E$) dependence of $\bar{p}/p$ ratio will show anomalous behavior~\cite{PhysRevLett.101.122302}. A detail calculation to demonstrate how the focusing effect could lead to anomalous $\beta_T$ dependence of $\bar{p}/p$ ratio has been done~\cite{LUO2009268}. It was found the $\bar{p}/p$ ratio will show different $\beta_T$ dependence trends with or without the QCP focusing effect. However, we did not observe this anomaly in $\beta_T$ dependence of $\bar{p}/p$ in Au+Au collisions at RHIC-BES measured by STAR experiment~\cite{PhysRevC.96.044904}. There are several reasons could suppress the focusing effect on $\bar{p}/p$. First, the contributions of strong and weak decay to proton and anti-proton are important in heavy-ion collisions~\cite{Yu:2018ijt,PhysRevC.92.064908,PhysRevC.93.054906}. Second, final state hadronic interactions between particles will dilute the QCP focusing effect. In this letter, we propose the transverse velocity dependence of $\bar{d}/d$ ratio or heavier light anti-nuclei to light nuclei ($^3\overline{\text{He}}/^3\text{He}$, $\bar{t}/t$, ......) ratios as more robust signatures of searching for the QCP.  {Assuming thermal production of the light nuclei along the system evolution trajectory, the yield ratio of light nuclei $\bar{d}/d$ is more sensitive to the $\mu_{B}$ than $\bar{p}/p$, due to the ratio $r \propto \exp[-2A\times\mu_{B}/T]$, $A$ is the mass number of the particle. It means the production of light nuclei is more sensitive to the system evolution trajectory in the vicinity of QCP, which will cause the changing of $T$ and $\mu_{B}$ of the system. One of another advantages is that the decay contributions for light nuclei is negligible in heavy-ion collisions.} In the following, we will formulate the QCP focusing effect on the $\beta_T$ dependence of $\bar{d}/d$ and $^3\overline{\text{He}}/^3\text{He}$ ratios by applying the UrQMD and THERMUS model to calculate the dependence patterns for a hypothetical focusing trajectory. 

\section{The QCD critical point focusing effect}

\begin{figure}
\includegraphics[width=0.48\textwidth]{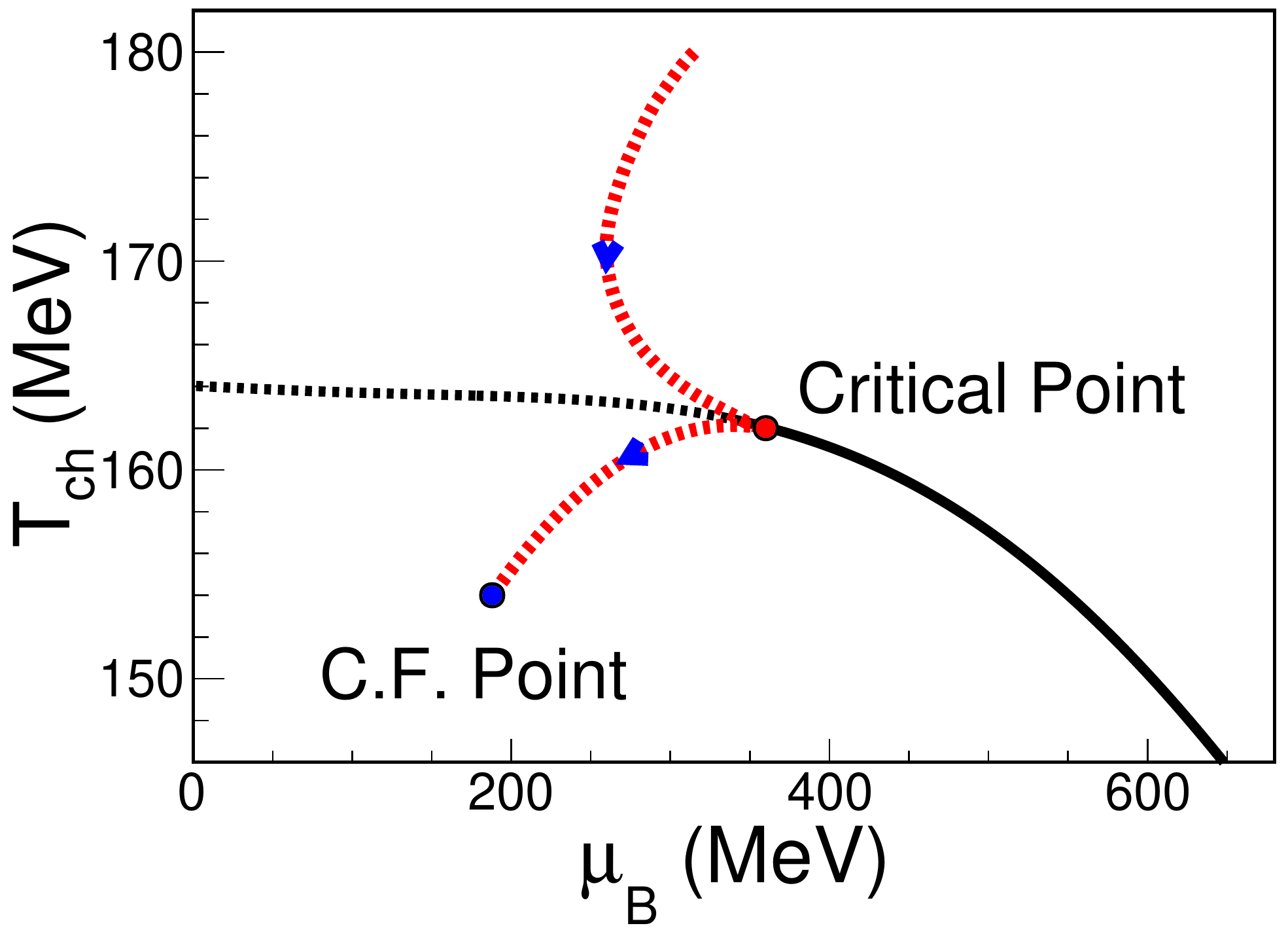}
\caption{\label{fig1} A sketch of conjectured QCD phase diagram with crossover (black dashed line), $1^{st}$ order phase transition boundary (black solid lines) and QCD critical point (red solid circle, $(T,\mu_B) = (162,360)$ MeV). A hypothetical system evolution trajectory (red dashed lines) is also plotted and ended with the chemical freeze-out point (blue solid circle).}
\end{figure}

In order to simulate the focusing effect, we assume that the critical point lies at $(T,\mu_B) = (162,360)$ MeV~\cite{1126-6708-2004-04-050} and the system evolution receives the focusing effect in central Au + Au collisions at $\snn = $19.6 GeV with chemical freeze-out point at $(T_{ch},\mu_B) = (152, 188)$ MeV~\cite{PhysRevC.96.044904}. Besides the starting (critical point) and ending (chemical freeze-out point) points, the hydrodynamic conjectured trajectory with focusing effect is shown in Fig.~\ref{fig1}. Follow the methods in Ref.~\cite{PhysRevC.71.044904,LUO2009268}, the normalized relative time $t = L/L_{\text{tot}}$ is used to characterize the time scale of the isentropic trajectory on the QCD phase diagram. The system is evolving from the critical point along the conjectured trajectory to the chemical freeze-out point. The $L$ represents the path length along the trajectory from the critical point to considered point and $L_{\text{tot}}$ is the total path length along the trajectory from the critical point to the chemical freeze-out point. The system is assumed to be thermodynamical equilibrium and is continuing to emit particles. Numbers of particle $A$ emitted at time $t$ along the trajectory is calculated by 
\begin{equation}
D_A(t) = \frac{Y_A[T(t),\mu_B(t)]}{\int_0^1 Y_A(t)dt}\times Y_A(t=1)
\end{equation}
where $A$ is the type of particle. $Y_A(t)$ is the yield of particle $A$ at a certain point on the trajectory, which is
determined by a statistical thermal model THERMUS~\cite{WHEATON200984}. $Y_A(t=1)$ is the yield at chemical freeze-out point and gives the normalization condition $\int_0^1 D_A(t)dt = Y(t=1)$.
{It means the sum of the total number of emitting particle $A$ equal to the particle multiplicity at chemical freeze-out. }

\begin{figure}
\includegraphics[width=0.48\textwidth]{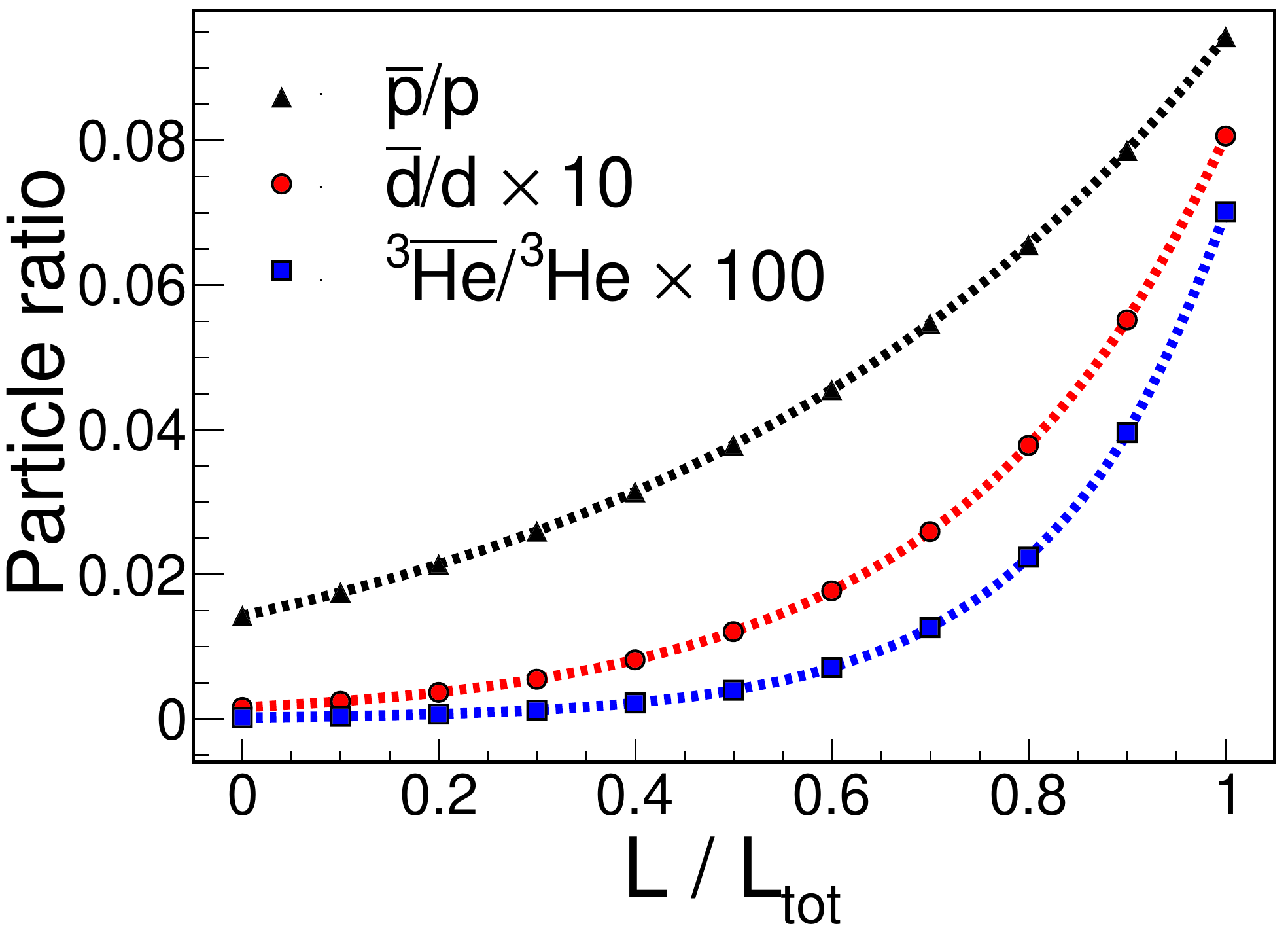}
\caption{\label{fig2} The time evolution of the $\bar{p}/p$, $\bar{d}/d$, and $^3\overline{\text{He}}/^3\text{He}$ ratios along the focusing trajectory are shown.}
\end{figure}

Time evolution of the particle ratios $N_{\bar{p}}(t)/N_p(t)$, $N_{\bar{d}}(t)/N_d(t)$, and $N_{^3\overline{\text{He}}}(t)/N_{^3\text{He}}(t)$ for the focusing effect trajectory are shown in Fig.~\ref{fig2}. Those ratios show an increasing trend as a function of time from the critical point ($t = 0$) to the chemical freeze-out point ($t = 1$) caused by the decreasing $\mu_B/T$ ratio along the focused trajectory. Due to the QCP focusing effect, the time evolution of three particle ratios is different and should be proportional to $\exp[-2A\times\mu_{B}/T]$. The $N_{\bar{d}}(t)/N_d(t)$ is more gradual at earlier stage and more abrupt at later stage than $N_{\bar{p}}(t)/N_p(t)$.

\begin{figure}
\includegraphics[width=0.48\textwidth]{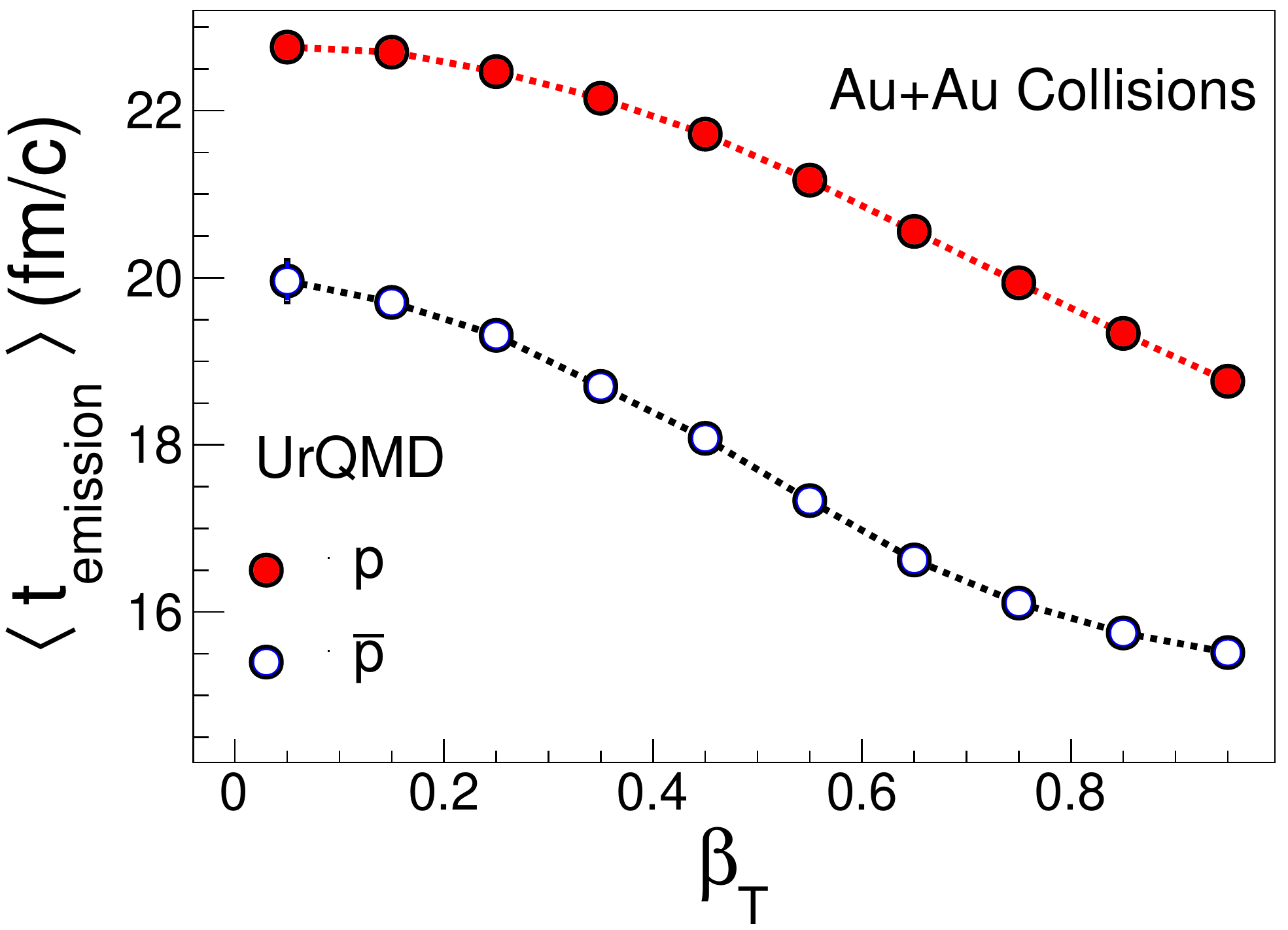}
\caption{\label{fig3} The UrQMD calculations for the $\beta_T$ dependence of average emission time of $p$ and $\bar{p}$ at mid-rapidity $|y| < 0.3$ in Au+Au collisions at $\snn = $ 19.6.}
\end{figure}

In order to obtain the $\beta_T$ dependence of those ratios, one needs to know the relation between emission time $t$ and transverse velocity $\beta_T$.  This relation can be obtained quantitatively by transport model, UrQMD~\cite{BASS1998255}. UrQMD is based on relativistic Boltzmann dynamics involving binary hadronic reactions, which are commonly used to describe the freeze-out and breakup of the fireball produced in relativistic heavy-ion collisions into hadrons. Two dimensions of $\beta_T - t$ distribution for $p$ and $\bar{p}$, $N_p(\beta,t)$ and $N_{\bar{p}}(\beta,t)$ are calculated by UrQMD Au+Au collisions at $\snn = $19.6 GeV with impact parameters $b <$ 4 fm. The average emission time $\langle t_{\text{emission}}\rangle$ as a function of $\beta_T$ of $p$ and $\bar{p}$ from UrQMD are shown in Fig.~\ref{fig3}. We observe strong $\beta_T - t$ anti-correlation for $p$ and $\bar{p}$ during the evolution of the system. It indicates the particles with larger transverse velocity are freeze-out at earlier time. We also found $\langle t_{\text{emission}}\rangle$ for $p$ are larger than $\bar{p}$ for a certain $\beta_{T}$, which suggests larger freeze-out time for protons than anti-protons.

\begin{figure}
\includegraphics[width=0.48\textwidth]{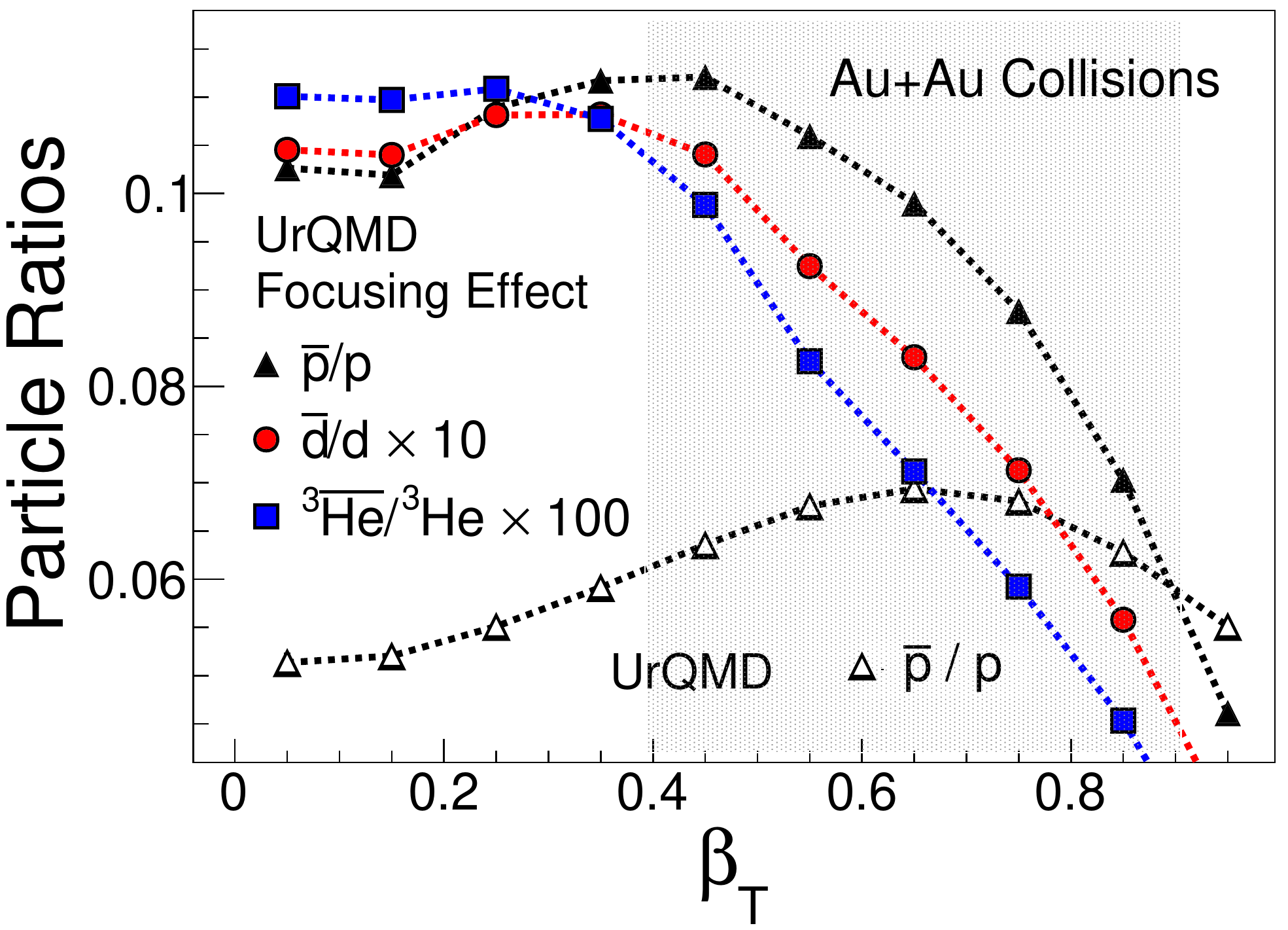}
\caption{\label{fig4}$\bar{p}/p$, $\bar{d}/d$, and $\bar{t}/t$ as a function of $\beta_T$ from UrQMD and UrQMD + QCP focusing effect. The band represents the range of $p_T / A$ from 0.5 to 2 GeV$/c$. $A$ is the mass number of light nuclei.}
\end{figure}

Once obtaining the relation between emission time $t$ and transverse velocity $\beta_T$, we can calculate the $\beta_T$ dependence of $\bar{p}/p$ ratio (solid triangles) as
\begin{equation}\label{ratios}
\frac{\bar{p}(\beta_T)}{p(\beta_T)}=\frac{\int N^{\text{U}}_{\bar{p}}(\beta_T,t)dt}{\int N^{\text{U}}_p(\beta_T,t)dt},
\end{equation}
where $N^{\text{U}}_{\bar{p}}(\beta_T,t)$ and $N^{\text{U}}_p(\beta_T,t)$ are the $\beta_T - t$ distribution for $\bar{p}$ and $p$, respectively. The results are shown in Fig.~\ref{fig4}. The $\bar{p}/p$ ratio from UrQMD shows an increasing trend as a function of $\beta_T$ (upto $\beta_T$=0.6) in the absence of QCP focusing effect, as the UrQMD does not include the physics of critical point. The $\bar{d}/d$ and $^3\overline{\text{He}}/^3\text{He}$ ratios from UrQMD should show similar trend as $\bar{p}/p$ ratio,  if the probability is similar of forming a light nuclei from nucleons and anti-nuclei from anti-nucleons. 

In order to obtain the $\beta_T$ dependence of anti-particle to particle ratio with QCP focusing effects, we convolute the time evolution of these ratios from Fig.~\ref{fig2} with the $\beta_T - t$ distribution from UrQMD. The multiplicity of certain particle with $\beta_T$ and $t$ is calculated by Thermal model. That means the $\beta_T - t$ distribution of particle with QCP focusing effect can be calculated by
\begin{equation} \label{QCP}
N^{\text{FE}}_{A}(\beta_T,t)=\frac{N^{\text{U}}_A(\beta_T,t)}{{\int N^{\text{U}}_A(\beta_T,t)}d\beta_T}\times D_{A}(t),
\end{equation}
where $A = \bar{p},p,\bar{d},d,...$. The normalized $\beta_T - t$ distribution $\dfrac{N^{\text{U}}(\beta_T,t)}{\int N^{\text{U}}(\beta_T,t)d\beta_T}$ for $d$ and $\bar{d}$ are assumed to be the same as those for $p$ and $\bar{p}$ in this study, as the light nuclei are coalesced by nucleons. The $\beta_T - t$ distribution of $^3\overline{\text{He}}$ and $^3\text{He}$ or heavier light nuclei can also be derived from the equations above. By using $\beta_T - t$ distribution of particles with QCP focusing effect obtained in Eq.~\ref{QCP}, the  $\beta_T$  dependence of anti-particle to particle ratio can be calculated by Eq.~\ref{ratios}. 

We show the $\beta_T$ dependence of $\bar{p}/p$, $\bar{d}/d$ and $^3\overline{\text{He}}/^3\text{He}$ ratios with QCP focusing effect in the Fig.~\ref{fig4}. The $\beta_T$ dependence of $\bar{t}/t$ (triton) is similar to the results of $^3\overline{\text{He}}/^3\text{He}$ due to the similar particle yield of the two types of particle. By comparing the $\bar{p}/p$ results from pure UrQMD calculations with those receiving QCP focusing effect, we find very different  $\beta_T$ dependence trends. It means the QCP focusing effect can lead to anomaly in $\beta_T$ dependence of anti-particle to particle ratio. We observed that the slope of these ratios are almost flat at low $\beta_T$ and become negative at higher $\beta_T$.  In our study, it shows that the heavier light nuclei is more sensitive to QCP.  The heavier the particle is, the steeper slope we can observe. However, the production for anti-light nuclei is difficult to be measured at lower collision energy~\cite{Adam:2019wnb}. Thus, we propose using $\beta_T$ dependence of anti-deuteron to deuteron ratio to search for QCD critical point in heavy-ion collisions. 

\begin{figure}
\includegraphics[width=0.48\textwidth]{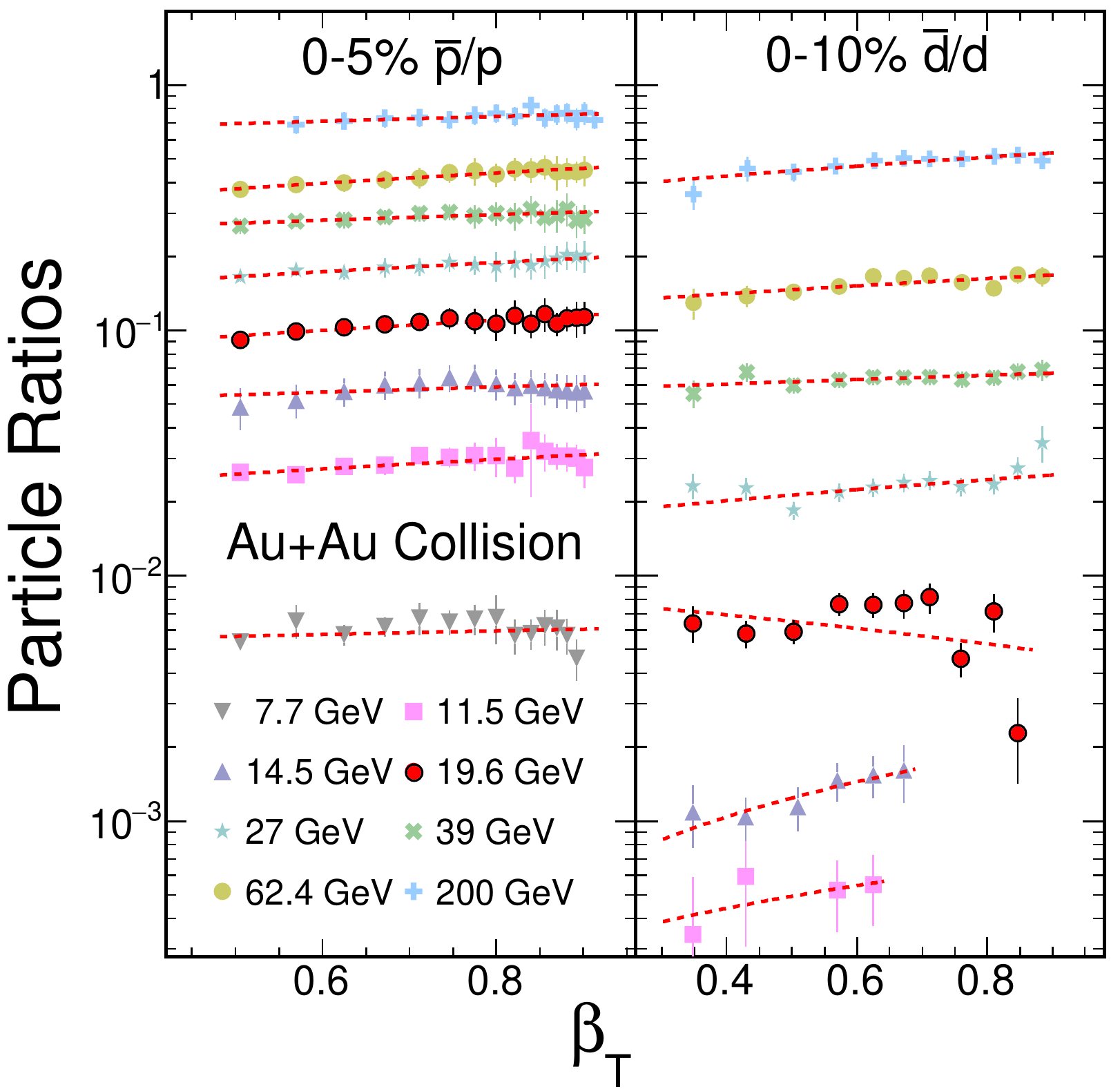}
\caption{\label{fig5} The $\beta_T$ dependence of 0-5\% central $\bar{p}/p$ (left) and 0-10\% central $\bar{d}/d$ (right) are derived from the $p_{T}$ spectra in Au+Au collisions measured by the STAR experiment at RHIC-BES energies~\cite{PhysRevLett.121.032301,Adam:2019wnb,YU2017788}. The dashed lines are linear fit. The error bars shown in the figure combine both of the systematic and statistical errors.}
\end{figure}

Experimentally, one needs to measure the $\beta_T$ dependence of $\bar{p}/p$ ratios as a function of energy, centrality and rapidity and do linear fits to obtain slopes. Negative slopes could indicate the system trajectories have passed through the critical region and the QCP is located on the right of the chemical freeze-out point of this collision energy due to the focusing effect. Then, a finer scan by looking at rapidity and centrality dependence of the slopes can further help to locate the QCP and the width of the critical region in the QCD phase diagram.  The $p_{T}$ spectra of $p(\bar{p})$ and $d(\bar{d})$ at mid-rapidity have been measured in Au+Au collisions by the STAR experiment at RHIC BES-I~\cite{PhysRevLett.121.032301,Adam:2019wnb,YU2017788,dingwei} with energies $\snn = $ 7.7-200 GeV. In Fig.~\ref{fig5}, the $\beta_T$ dependence of 0-5\% collision centrality for $\bar{p}/p$ and 0-10\% for $\bar{d}/d$ ratios are shown. The longitudinal momentum $p_z$ is smaller than the energy of particle at mid-rapidity, the approximation $\beta_T=p_T/E\approx p_T/\sqrt{m_0^2+p^2_T}$ is used in our analysis, where $E$ and $m_0$ are the energy and mass of particle. We did linear fits to these data and found positive slopes for $\beta_T$ dependence of $\bar{p}/p$. The positive slopes for $\beta_T$ dependence of $\bar{d}/d$ are also observed for all energies except 19.6 GeV. The decreasing trend of $\bar{d}/d$ at high $\beta_T$ in central Au+Au collisions at $\snn = $19.6 GeV is consistent with the trend in Fig.~\ref{fig4} with QCP focusing effect. If the anomaly in $\beta_T$ dependence of $\bar{d}/d$  at 19.6 GeV is indeed due to the QCP focusing effect, it indicates the system evolution trajectories have passed through the critical region and the $\mu_B$ of the QCP should be larger than the chemical freeze-out $\mu_B$ of 19.6 GeV.  Currently, we observe a positive slope for the $\beta_T$ dependence of $\bar{d}/d$ at 14.5 and 11.5 GeV. However, this could be due to the limited statistics, which makes it difficult to measure the high $\beta_T$ region, especially for $\bar{d}$. 

\section{Summary}
We studied the QCP focusing effect on $\beta_T$ dependence of $\bar{p}/p$, $\bar{d}/d$, and $^3\overline{\text{He}}/^3\text{He}$ ratios. The focusing effect is modeled by convoluting the particle density along the focused trajectories and the $\beta_T - t$ distribution from UrQMD model. The focusing effect will lead to a decreasing anti-particle to particle ratio when increasing $\beta_T$. We examined and did a linear fit to the $\beta_T$ dependence of $\bar{p}/p$ and $\bar{d}/d$, which are calculated from the STAR measured $p_{T}$ spectra. We observed that only the fitting slope of the $\bar{d}/d$ at $\snn = $19.6 GeV is negative. The negative slope can be qualitatively explained in term of the QCP focusing effect, which might indicate the system evolution trajectory at $\snn = $19.6 GeV has passed through the critical region. This anomaly could be potentially connected with the dip and peak structures observed at 19.6 GeV in the measurements of net-proton fluctuations and neutron density fluctuations by STAR experiment, respectively.  
We can make more precise measurements and further constraint on the $\mu_B$ value of QCP in the second phase of Beam Energy Scan program (BES-II, 2019-2021) at RHIC~\cite{Bzdak:2019pkr}.  Furthermore, since $\mu_B$ depends on rapidity, we could also do rapidity scan for $p_T$ dependence of $\bar{d}/d$ at each energy. This might allow us to map out the location of the QCP with finer $\mu_B$ step. Finally, we predicted the $\beta_T$ dependence of heavier anti-light nuclei to light nuclei ratio, such as $^3\overline{\textrm{He}} / ^3 \textrm{He}$ and $\bar{t}/t$, are more sensitive to the QCP focusing effect.

\section{Acknowledgement}
We thank Dr. Nu Xu for the fruitful discussions. This work is supported in part by the National Natural Science Foundation of China under Grants 
(No. 11890711,11575069, 11828501 and 11861131009), Fundamental Research Funds for the Central Universities No. CCNU19QN054, , Nanhu Scholar Program for Young Scholars of XYNU  and CCNU-QLPL Innovation Fund (Grant No. QLPL201801).
%\tableofcontents
\bibliography{CEP}
\end{document}